\documentclass[twocolumn,showpacs,pra,aps,amssymb]{revtex4}
\begin{document}
\title{\bf Steady state entanglement of two atoms created by classical driving field}

\author{ \"{O}zg\"{u}r \c{C}akir, Alexander A. Klyachko and Alexander S. Shumovsky}

\affiliation{Faculty of Science, Bilkent University, Bilkent,
Ankara, 06800 Turkey}

\begin{abstract}
The stabilization of steady state entanglement caused by action of
a classical driving field in the system of two-level atoms with
the dipole interaction accompanied by spontaneous emission is
discussed. An exact solution shows that the maximum amount of
concurrence that can be achieved in Lamb-Dicke limit is $0.43$,
which corresponds to the entanglement $\mathcal{E}_{max}=0.285$
ebit. Dependence of entanglement on interatomic distance and
classical driving field is examined numerically.
\end{abstract}

\pacs{PACS numbers: 03.65.Ud, 32.80.-t, 42.50.Ct}

\maketitle

\section{Introduction}

It is well known that quantum entanglement plays fundamental role
in the quantum information processing and quantum computing
\cite{1,2}. The practical application of entanglement requires the
{\it robust} entangled states. This notion includes long enough
lifetime of the states and high amount of entanglement (as close
to maximum entanglement as possible).

The simplest entangled states of two or more qubits can be
modelled physically by the states of two-level atoms, which have
been successfully used for decades as the main tool for testing
fundamentals of quantum mechanics \cite{3}. In particular, a
number of important experiments on generation and manipulation
with entangled states in atomic systems have been performed
recently (e.g., see \cite{4} and references therein).

The lifetime of entangled states in atomic systems is usually
governed by the spontaneous emission rate and therefore is quite
short \cite{5}. The problem of generation of robust entangled
states in atomic systems is of high importance. In fact, this
problem belongs to a more general class of problems related to the
so-called {\it quantum decoherence}. This branch of modern physics
examines an irreversible evolution of quantum systems under
influence of a dissipative environment (see \cite{6} and
references therein). Thus, using the ideas of quantum decoherence
theory, we have to choose a proper environment for an atomic
system, providing an irreversible evolution of this system towards
a stable (or at least metastable) entangled state.

For example, it has been proposed in Refs. \cite{7,8} to use the
three-level atoms with $\Lambda$-type configuration of levels
instead of the two-level atoms. In this case, an unstable maximum
entangled state is created by absorption of a photon, resonant
with the transition between the ground and highest excited level
in the system of the two atoms. This state rapidly decays either
to the ground state or to the intermediate excited state. The
latter process is accompanied by generation of the so-called
Stokes photon and leads to creation of a new maximum entangled
state of two atoms. If the Stokes photon, generated by the
transition between the highest and intermediate excited states, is
then discarded through the use of an appropriate dissipative
environment, the system remains in the maximum entangled state for
a long time determined by the non-radiative decay of the
intermediate atomic state.

In the above example, the maximum entangled atomic state is
induced by the photon exchange between the atoms. Another physical
proposal is based on the use of direct dipole interaction between
the two-level atoms \cite{9}. In this case, the energy loss in the
system due to the spontaneous emission is supposed to be
compensate by interaction with an environment. In Refs. \cite{9}
it was proposed to use the interaction of atoms with the squeezed
vacuum state in addition to the dipole-dipole interaction. It was
shown that a steady state with reasonable amount of entanglement
can be achieved in this case.

This is not an unexpected result. The point is that it has been
recognized recently that entanglement is peculiarly connected with
the amount of quantum fluctuations in the system
\cite{7,10,11,12}.  This fact can be used as an operational
definition of entanglement \cite{7,12}.

By the total amount of quantum fluctuations we mean the sum of
variances of observables $\mathcal{O}_i$ which form an orthonormal
basis of Lie algebra of the dynamic symmetry group $G$ of quantum
system \cite{11}. For example. in the case of two qubits (two
two-level atoms, the dynamic symmetry group is $G=SU(2) \times
SU(2)$, while the local observables are provided by the Pauli
(spin projection) operators.

The environment of Ref. \cite{9} provided by the squeezed vacuum
always manifests a high level of quantum fluctuations, and by
means of interaction with the atomic system may transmit the
quantum fluctuations to the atomic system, which is supposed to be
initially unentangled.

It seems to be interesting to consider the case when an entangled
state is generated from initially unentangled state in the
presence of non-fluctuating classical environment. In this note,
we show that a reasonable amount of entanglement can be achieved
in a system of two-level atoms with dipole interaction in presence
of a classical driving field. Although the classical driving field
is devoid of quantum fluctuations, it helps to realize the
potential ability of entanglement hidden in the structure of the
eigenstates of the system. Let us note that the stabilizing
influence of the classical driving field on entanglement in a
system of two-level atoms in a cavity with high quality factor has
been discussed in Ref. \cite{13}.

The paper is arranged as follows. In Sec. II, we discuss the model
Hamiltonian and its properties. We examine the structure of
eigenstates and show the presence of entanglement. We show that
basis of the corresponding Hilbert space can be divided into the
antisymmetric singlet part, represented by a certain maximum
entangled state, and the triplet part. In Sec. III, we show that
under certain physical assumptions the master equation can be
solved in the triplet domain of the Hilbert space. Under the
assumption of short interatomic distances (Lamb-Dicke limit), we
find an exact steady state solution which give high enough amount
of entanglement. We also examine entanglement as a function of
interatomic separation and classical driving field.  Sec. IV
contains discussion of the obtained results and conclusions.

\section{Properties of the model Hamiltonian}

The two-atom system under consideration can be specified by the
Hamiltonian
\begin{eqnarray}
H= \sum_{i=1}^2 \left[
\frac{\Delta^{(i)}}{2}\sigma_z^{(i)}+E^{(i)} (
\sigma^{(i)}_+e^{i\vec{k}_0 \cdot \vec{r}} + \sigma^{(i)}_-e^{-i\vec{k}_0 \cdot \vec{r}}) \right] \nonumber \\
+\Omega(\sigma^{(1)}_+ \sigma^{(2)}_-+ \sigma^{(1)}_-
\sigma^{(1)}_+) . \label{1}
\end{eqnarray}
Here $\Delta^{(i)}= \omega^{(i)} - \omega_C$ is the detuning
between the atomic frequency $\omega^{(i)}$ and classical driving
field frequency $\omega_C$ assumed to be small ($\omega^{(i)} \gg
| \Delta^{(i)}|$), $E^{(i)}$ represents the classical driving
field at the location of the $i$-th atom, and the dipole-dipole
coupling constant has the form \cite{9}
\begin{eqnarray}
\Omega = \frac{3}{4} \sqrt{\Gamma^{(1)} \Gamma^{(2)}}
\left[-(1-|\vec{\mu} \cdot \vec{r}|^2) \frac{\cos k_0r}{k_0r}\right.\nonumber
\\\left. +(1-3|\vec{\mu} \cdot \vec{r}|^2) \left( \frac{\sin
k_0r}{(k_0r)^2} + \frac{\cos k_0r}{(k_0r)^3} \right)
\right],
\label{2}
\end{eqnarray}
where $\vec{\mu}^{(i)}$ is the atomic dipole moment,
$k_0=(\omega^{(1)}+\omega^{(2)})/2c$, and
\begin{eqnarray}
\Gamma^{(i)}= \frac{k_0^3 |\vec{\mu}^{(i)}|^2}{3 \pi \hbar
\epsilon_0}. \nonumber
\end{eqnarray}
This Hamiltonian (1) is defined in the four-dimensional Hilbert
space of two qubits $\mathcal{H}_{2,2}$ spanned by the vectors
\begin{eqnarray}
|gg \rangle , \quad |eg \rangle, \quad |ge \rangle , \quad |ee
\rangle , \label{3}
\end{eqnarray}
where $|\alpha \beta \rangle = |\alpha \rangle \otimes |\beta
\rangle$ and $\alpha = e,g$ denotes the excited and ground atomic
states, respectively. This space $\mathcal{H}_{2,2}$ represents in
fact a product of the two two-dimensional ``spin-$\frac{1}{2}$"
spaces $\mathcal{H}_{1/2}$ spanned by the vectors $|e \rangle$ and
$g
\rangle$ each. This space can also be expanded as follows
\begin{eqnarray}
\mathcal{H}_{2,2}=\mathcal{H}_{1/2} \otimes
\mathcal{H}_{1/2}=\mathcal{H}_0 \oplus \mathcal{H}_1 , \label{4}
\end{eqnarray}
where $\mathbb{H}_0$ is the singlet space, consisting of the
anti-symmetric maximum entangled state
\begin{eqnarray}
|A \rangle = \frac{1}{\sqrt{2}} (|eg \rangle - |ge \rangle )
\label{5}
\end{eqnarray}
(``spin-$0$" space). Using the terminology of the phase-transition
theory \cite{14}, we can  associate this state with the
``aniferromagnetic" order in the system of two dipoles.

In turn, the ``spin-$1$" space $\mathcal{H}_1$ is spanned by the
triplet of symmetric states
\begin{eqnarray}
\left\{ \begin{array}{lcl} |+1 \rangle & = & |ee \rangle \\ |0
\rangle & = & \frac{1}{\sqrt{2}} (|eg \rangle +|ge \rangle ) \\
|-1 \rangle & = & |gg \rangle \end{array} \right. \label{6}
\end{eqnarray}
Formally, this basis (6) can be associated with the states of a
``spin-1" object. Namely, the states $|+1 \rangle$, $|0 \rangle$,
and $|-1 \rangle$ correspond to the states with projections of
``spin" equal to $1$, $0$, and $-1$, respectively. It should be
noted that the state $|0 \rangle$ of a qutrit always manifests
maximum entanglement \cite{11,12}.

Using the analogy with the phase transition theory, we can say
that the unentangled states $|+1 \rangle$ and $|=1 \rangle$
correspond to the ``ferromagnetic" (parallel) order of the
dipoles, while the maximum entangled state $|0 \rangle$ in (6)
corresponds to the disordered ``paramagnetic" state.

Assume now that the atoms are identical when we may put
$\Delta^{(1)}=\Delta^{(2)} \equiv \Delta$ and
$\Gamma^{(1)}=\Gamma^{(2)} \equiv \Gamma$. In addition, we assume
that the polarization of classical driving field is parallel to
the interatomic axis, so that $E^{(1)}=E^{(2)}=E$. It can be
easily seen that under these conditions the antisymmetric singlet
state (5) is an eigenstate of the Hamiltonian (1) with the
eigenvalue $\epsilon_A=- \Omega$:
\begin{eqnarray}
H|A \rangle = - \Omega |A \rangle . \label{7}
\end{eqnarray}

In the triplet sector, equation for eigenvalues takes the form
\begin{eqnarray}
\epsilon^3-\Omega \epsilon^2 -( \Delta^2 +4E^2) \epsilon +\Delta^2
\Omega = 0. \label{8}
\end{eqnarray}
In the absence of external field ($E=0$), this equation has the
roots $\epsilon_{+1} = \Delta$, $\epsilon_0 = \Omega$, and
$\epsilon_{-1} =- \Delta$, and the set of the corresponding
eigenstates is provided by the states (6). Thus, at $\Omega >|
\Delta |$, the ``antiferromagnetic" maximum entangled state (5) is
the ground state of the system in the absence of the external
field. Otherwise, one of the states with ``ferromagnetic"
(parallel) order of atomic dipoles becomes the ground state. This
is quite natural result, because the pseudo-spin Hamiltonian (1)
has the same structure as the model of the so-called plane rotator
in transverse field, examined in the theory of phase transitions
\cite{14}. It is clear that similar behavior takes place at small
values of the classical driving field.

In the absence of detuning $(\Delta =0)$ and an arbitrary value of
the classical driving field, there is the following symmetric
maximum entangled eigenstate
\begin{eqnarray}
|\psi_0 \rangle = \frac{1}{\sqrt{2}} (|ee \rangle - |gg \rangle)
\nonumber
\end{eqnarray}
in the triplet sector, which corresponds to the eigenvalue
${\varepsilon}_0=0$, and the two eigenstates
\begin{eqnarray}
|\psi_{\pm} \rangle = \frac{1}{\sqrt{4E^2+ {\varepsilon}_{\pm}^2}}
[E \sqrt{2} (|ee \rangle +|gg \rangle ) +{\varepsilon}_{\pm} |0
\rangle ] \label{9}
\end{eqnarray}
with the eigenvalues
\begin{eqnarray}
{\varepsilon}_{\pm}= \frac{\Omega}{2} \pm \frac{1}{2}
\sqrt{\Omega^2+16E^2} . \label{10}
\end{eqnarray}
The concurrence of the states (9) is then given by the expression
\cite{12}
\begin{eqnarray}
C=2|\det [\psi_{\pm}]|=
\frac{|4E^2-\varepsilon_{\pm}^2|}{4E^2+\varepsilon^2_{\pm}} ,
\label{11}
\end{eqnarray}
where $[\psi_{\pm}]$ denotes the matrix of coefficients in the
states (9). For the lowest state $|\psi_- \rangle$ with the
eigenvalue $\varepsilon_-$ in (10), the concurrence (11) achieves
the maximum value $C=1$ at $E=0$ and then monotonically decreases
with the increase of $E$. Similar behavior takes place in the case
of state $|\psi_+\rangle$ with the eigenvalue $\varepsilon_+$.

Thus, the eigenstates of the system with the Hamiltonian (1) can
manifests entanglement and even maximum entanglement (at least at
a certain choice of parameters). Therefore, all one can assume is
that the system under consideration has a good potential ability
to evolve towards an entangled state under influence of a proper
dissipative environment.

\section{Irreversible evolution towards entanglement}

The irreversible evolution assumes dissipation of energy. In the
system under consideration, it can be described by the losses due
to spontaneous decay of the excited atomic states. We can specify
the irreversible evolution in the system described by the
Hamiltonian (1) with the assumptions made in the previous Section
by the following master equation \cite{16}
\begin{eqnarray}
\dot{\rho}=-i[H, \rho ]+ \frac{1}{2} \sum_{i,j=1}^2 \Gamma^{(ij)}
(2 \sigma_-^{(i)} \rho \sigma_+^{(j)} \nonumber \\ -\sigma_+^{(i)}
\sigma_-^{(j)} \rho - \rho \sigma_+^{(i)} \sigma_-^{(j)} ),
\label{12}
\end{eqnarray}
where the coefficients $\Gamma^{(ij)}$ are responsible for the
spontaneous decay and have the form $\Gamma^{(ii)}=\Gamma^{(i)}$
and
\begin{eqnarray}
\Gamma^{(12)}= \Gamma^{(21)} =-3 \sqrt{\Gamma^{(1)} \Gamma^{(2)}}
\ \left[ \frac{\cos k_0r}{(k_0r)^2} - \frac{\sin k_0r}{(k_0r)^3}
\right] . \label{13}
\end{eqnarray}

We now show that under a certain choice of the initial state, the
antiferromagnetic maximum entangled state (5) can be discarded
from the evolution picture described by the master equation (12).
Since (5) is the eigenstate of the Hamiltonian (1), it does not
contribute into the unitary evolution, described by the first term
in the right-hand side of (12).  We also have
\begin{eqnarray}
\sum_{i=1}^2 \sigma_-^{(i)}|A\rangle =0. \nonumber
\end{eqnarray}
Thus, it does not contribute into the Liouville term in the
right-hand side of (12). This means that if the system is prepared
initially in the maximum entangled antisymmetric state (5), it
will stay in this state forever. If the initial state is chosen
differently, the irreversible evolution of the system should be
considered in the triplet sector (6).

This means that the density matrix in Eq. (12) takes the following
block form
\begin{eqnarray}
\rho = \left( \begin{array}{cccc} \rho_{11} & \rho_{12} &
\rho_{13} & 0 \\ \rho_{21} & \rho_{22} & \rho_{23} & 0 \\
\rho_{31} & \rho_{32} & \rho_{33} & 0 \\ 0 & 0 & 0 & \rho_{44}
\end{array} \right) \label{14}
\end{eqnarray}
Here  the blocks correspond to the bases (6) and (5),
respectively. The non-Hermitian effective Hamiltonian,
corresponding to the evolution described by the master equation
(12) with the density matrix (14), has the form
\begin{widetext}
\begin{eqnarray}
H_{eff} = \left( \begin{array}{cccc} \Delta -i \Gamma & \sqrt{2} E
& 0 & 0 \\ \sqrt{2} E & \Omega - \frac{i}{2} (\Gamma +
\Gamma^{(12)} ) & \sqrt{2} E & 0 \\ 0 & \sqrt{2} E & - \Delta  & 0
\\ 0 & 0 & 0 & - \Omega - \frac{i}{2} ( \Gamma - \Gamma^{(12)} )
\end{array}
\right) \label{15}
\end{eqnarray}
\end{widetext}
Here again $\Gamma^{(1)}=\Gamma^{(2)} = \Gamma$. It is seen from
the non-Hermitian Hamiltonian (15) that in the absence of the
classical driving field, all states except $|gg \rangle$ are
damped, so that the steady state entanglement at $E=0$ is
impossible, and the system evolves towards the unentangled state
$|gg \rangle$.

In other words, only the presence of classical driving field can
stabilize steady state entanglement in the model system with the
Hamiltonian (1).

Because we are interested in the robust entanglement, let us
consider the steady state solutions of the master equation (12)
with the density matrix (14) in the triplet sector of the Hilbert
space.

Consider first the Lamb-Dicke limit of short interatomic
separation. Then, it follows from the definition of the decay rate
(13) that
\begin{eqnarray}
\Gamma^{(12)}=\Gamma^{(21)} \approx \Gamma . \nonumber
\end{eqnarray}
In this case, making the further assumption that $\Delta =0$, for
the steady state density matrix in triplet sector we get the
following explicit expression
\begin{widetext}
\begin{eqnarray}
\rho_S= \frac{1}{N} \left( \begin{array}{ccc} 64 E^4 & -16iE^3
\sqrt{2} & 8E^2(2i \Omega-1) \\ 16i E^3 \sqrt{2} & 8E^2 (1+8E^2) &
-2E\sqrt{2}(2 \Omega+i+8i E^2) \\ -8E^2(2i \Omega+1) &
-2E\sqrt{2}(2 \Omega-i-8i E^2) & 4(\Omega^2 +2E^2+16E^4)+1
\end{array} \right) \label{16}
\end{eqnarray}
\end{widetext}
Here $N=Tr \rho$ is the normalization factor  and $\Omega$ and $E$
are established for the dimensionless parameters $\Omega /\Gamma$
and $E/\Gamma$, respectively. It follows from the numerical
analysis that the detuning  $\Delta$  has a weak influence upon
the concurrence.

To determine the settings, leading to the maximum possible amount
of entanglement in the system under consideration, we choose
$\Omega = \tau E^2$,  where $\tau$ is a dimensionless constant to
be determined upon the maximization of concurrence. This factor in
the Lamb-Dicke limit can be represented as follows
\begin{eqnarray}
\tau = \frac{3}{4 \pi \alpha} [(k_0r)^3 Q \bar{n}V]^{-1},
\label{17}
\end{eqnarray}
where $\alpha = 1/137$ is the fine structure constant, $Q$ denotes
atomic quality factor ($Q= \omega_0 T$, where $T$ is the lifetime
of the excited atomic state), $\bar{n}$ is the mean number of
photons per unit volume in classical driving field, and $V$
denotes the volume of interaction between atom and field, so that
$\bar{n}V$ gives the mean number of photons interacting with atom
during the time $T$.

The concurrence is defined as follows \cite{15}
\begin{eqnarray}
C= \max (\lambda_1-\lambda_2 -\lambda_3,0), \label{18}
\end{eqnarray}
where $\lambda$ denotes the spectrum of matrix $R=( \sqrt{\rho}
\bar{\rho} \sqrt{\rho} )^{1/2}$ and $\bar{\rho}$ denotes the
complex conjugation of (16) in the so-called ``magic basis"
\cite{15}. The maximum entangled state provides $C=1$, while the
unentangled states give $C=0$.

One can see from Eq. (2) that at fixed $\tau$ and in the
Lamb-Dicke limit $\vec{k}_0 \cdot \vec{r} \ll 1$, both
dimensionless parameters $\Omega / \Gamma , E/ \Gamma \gg 1$. In
this case, the density matrix (16) takes the form
\begin{eqnarray}
\rho_S \approx \frac{1}{\tau^2+48} \left( \begin{array}{ccc} 16 & 0 & 4i \tau \\
0 & 16 & 0 \\ -4i \tau & 0 & 16 + \tau^2 \end{array} \right)
\label{19}
\end{eqnarray}
To our surprise, the concurrence (18) in this limit turns out to
be rational function of $\tau$
\begin{eqnarray}
C(\tau) = \frac{8 \tau -16}{\tau^2+48},\quad \tau\ge 2\nonumber
\end{eqnarray}
extended by zero at $\tau \leq 2$. Thus, entanglement is
impossible if $\tau \leq 2$. The maximum value of concurrence
\begin{eqnarray}
C_{max}= \frac{2}{\sqrt{13}+1} \approx 0.43 \nonumber
\end{eqnarray}
is attained at
\begin{eqnarray}
\tau_{max} = 2+2\sqrt{13} \approx 9.21. \nonumber
\end{eqnarray}
The corresponding amount of entanglement \cite{15} is
\begin{eqnarray}
\mathcal{E}_{max}=H \left( \frac{1- \sqrt{1-C_{max}^2}}{2} \right)
\approx 0.285 \quad \mbox{ebit.} \nonumber
\end{eqnarray}
Taking into account the form of the dimensionless parameter $\tau$
given by Eq. (17), we can examine the dimensionless interatomic
distance $\vec{k}_0 \cdot \vec{r}$, corresponding to the maximum
entanglement provided by $\tau_{max}=9.21$, as a function of the
number of photons $\bar{n}V$, which should obey the condition
$\bar{n}V \gg 1$ in the case of classical driving field. The
dependence is given by Fig. 1. It is seen that in the case of mean
number of photons $\bar{n}V \sim 10$, the interatomic distance
should be of the order of $10^{-2} \lambda$ (where $\lambda$ is
the wavelength) to achieve the maximum possible amount of
entanglement. Increase of the mean number of photons in the
driving field, considered as a coherent state with $|\alpha|^2 \gg
1$, leads to a decrease of interatomic distance, which is required
to have maximum amount of entanglement.

The dependence of concurrence from the interatomic distance comes
from the coupling constant $\Omega$ (2). The results of numerical
calculations for different interatomic separations and different
values of the classical driving field are shown in Fig. 2. It is
seen that  deviation from the Lamb-Dicke limit decreases the
concurrence.

Thus, the classical driving field stabilizes entanglement in the
system under consideration.

We now turn to the discussion of the obtained results.

\section{Summary and conclusion}

We have examined the system of two identical two-level atoms,
interacting with each other by means of dipole forces. The
dissipation of energy in the system is provided by the spontaneous
decay of the excited atomic states. The compensation of losses is
provided by the presence of classical driving field.

The consideration of the eigenstates of the Hamiltonian (1),
describing the system under consideration, shows that the system
carries a potential ability of creation of entanglement, hidden in
the structure of the eigenstates.

Instead of conventional picture of two-qubit system, we prefer to
specify the system by the triplet of states of an effective qutrit
and by an ``antiferromagnetic", antisymmetric maximum entangled
state, corresponding to a ``spinless" quasiparticle.

It is shown that in the absence of the classical driving field,
system evolves towards an unentangled state (both atomic dipoles
are in the ground state). The presence of the classical driving
field stabilizes the entanglement.

In the Lamb-Dicke limit of a point-like system, we obtained an
exact solution for the steady state density matrix, that manifests
high amount of entanglement (the concurrence $C_{max}=0.43$ and
the entanglement $\mathcal{E}_{max}=0.285$ ebit). This amount is
much higher than in a number of recent proposals. In particular,
it is higher than that in the case when the squeezed vacuum is
used for stabilization of entanglement instead of the classical
driving field \cite{9}. It should be stressed that this amount of
concurrence corresponds to a reasonable magnitude of the parameter
$E$, describing interaction of classical driving field with atoms.

Since the classical driving field is devoid of quantum
fluctuations, the only source of the high-level quantum
fluctuations peculiar for entanglement can be associated with the
structure of the eigenstates of the Hamiltonian (1), describing
the system. Thus, the classical driving field plays a double role.
Namely, it compensates the losses of energy and stabilizes the
entanglement carried by the system.

Such a picture suggests an analogy with the phase transitions in
antiferromagnets, whose ground state can be considered as a
collective entangled state. This ground state of such a system is
realized at low temperatures. In other words, it is stabilized by
the presence of a low-temperature environment (thermostat).

In our case, the system under consideration has very few degrees
of freedom and therefore cannot manifest a thermodynamic phase
transition. Nevertheless, the classical driving field together
with the spontaneous decay plays a role similar to that of the
thermostat.

In the investigation of steady state entanglement in the system,
we always considered  the triplet sector of the Hilbert space.
From the physical point of view, this means that the initial
states are symmetric. For example, the most natural choice is
provided by the unentangled initial state $|gg \rangle$ with both
atoms in the ground state. In this case, a certain balance between
the spontaneous decay and influence of the classical driving field
lead to creation of the steady state entanglement in the system.

Let us now discuss the effect caused by the inclusion of the
maximum entangled antisymmetric state (5) into the picture. This
means that we choose the density matrix of the form
$$
p|A\rangle \langle A| +q \rho_S,
$$
in the whole four-dimensional Hilbert space. Here $\rho_S$ denotes
the block, coming from the triplet sector and $p,q$ are the
weights of the corresponding contributions $(p+q=1)$. Since the
state $|A \rangle$ belongs to the ''magic basis" \cite{15}, the
concurrence of the steady state is given by formula similar to
(17) applied to spectrum $q \lambda_1 , q \lambda_2 , q \lambda_3,
p$. As far as $p<q \lambda_1$, that is $p< \lambda_1 / (1+
\lambda_1)$, the presence of the antisymmetric component (5) {\it
decreases} the concurrence.

Thus, to achieve a reasonable amount of robust entanglement, we
have to prepare the system initially either in the antisymmetric
maximum entangled state (5) or in a state from the symmetric
sector (6) of the Hilbert space.

In the above consideration, we always assumed that atoms are
identical. It seems to be interesting to extend our consideration
on the case of non-identical atoms. In view of the result of Ref.
\cite{9}, we can expect that this may lead to a significant
increase of amount of entanglement.

We also restricted our consideration to the case of polarization
of the classical driving field parallel to the interatomic axis.
The alternative choice of the polarization perpendicular to the
interatomic axis can lead to a strong changes of picture as well.
First of all, the change of polarization changes the form of the
coupling constant (2). Then, it causes the consideration of the
different values of the classical driving field in the atomic
locations. Finally, in this case, the antisymmetric state (5)
cannot be discarded at all, so that the irreversible evolution
should be examined in the whole four-dimensional Hilbert space.

The detailed analysis of the above mentioned two extensions of the
model deserves special consideration.

\newpage
\begin{figure}
\caption{Lamb-Dicke parameter $(\vec{k}_0 \cdot \vec{r})$ versus
the number of photons in the case of dimensionless parameter (17),
providing the maximal amount of entanglement for typical atomic
quality factor $Q=10^{-6}$.}
\end{figure}

\begin{figure}
\caption{Numerical dependence of concurrence on the interatomic
distance and classical driving field. The dimensionless quantities
$\Omega/\Gamma$ and $E/\Gamma$ are used here.}
\end{figure}

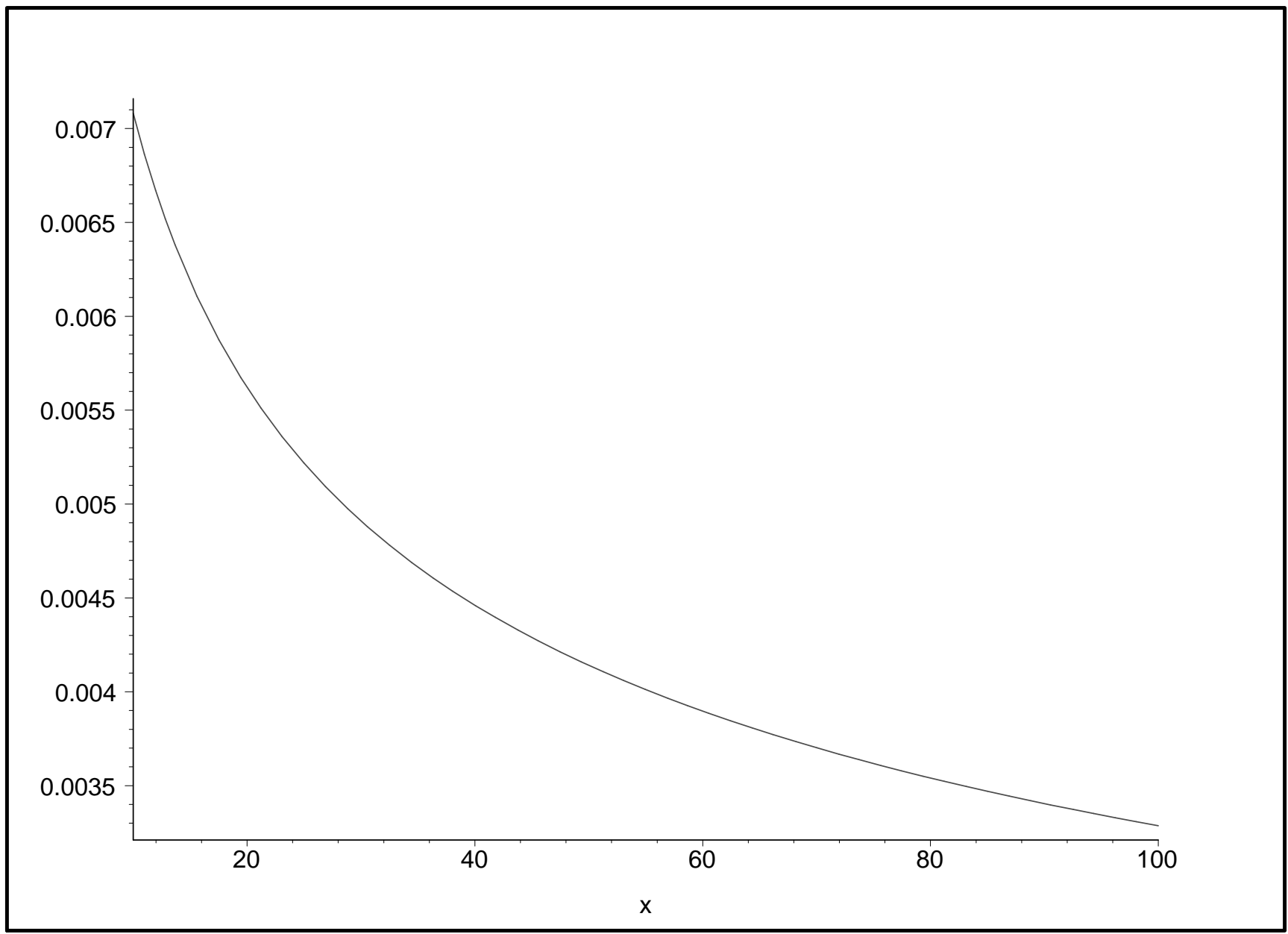
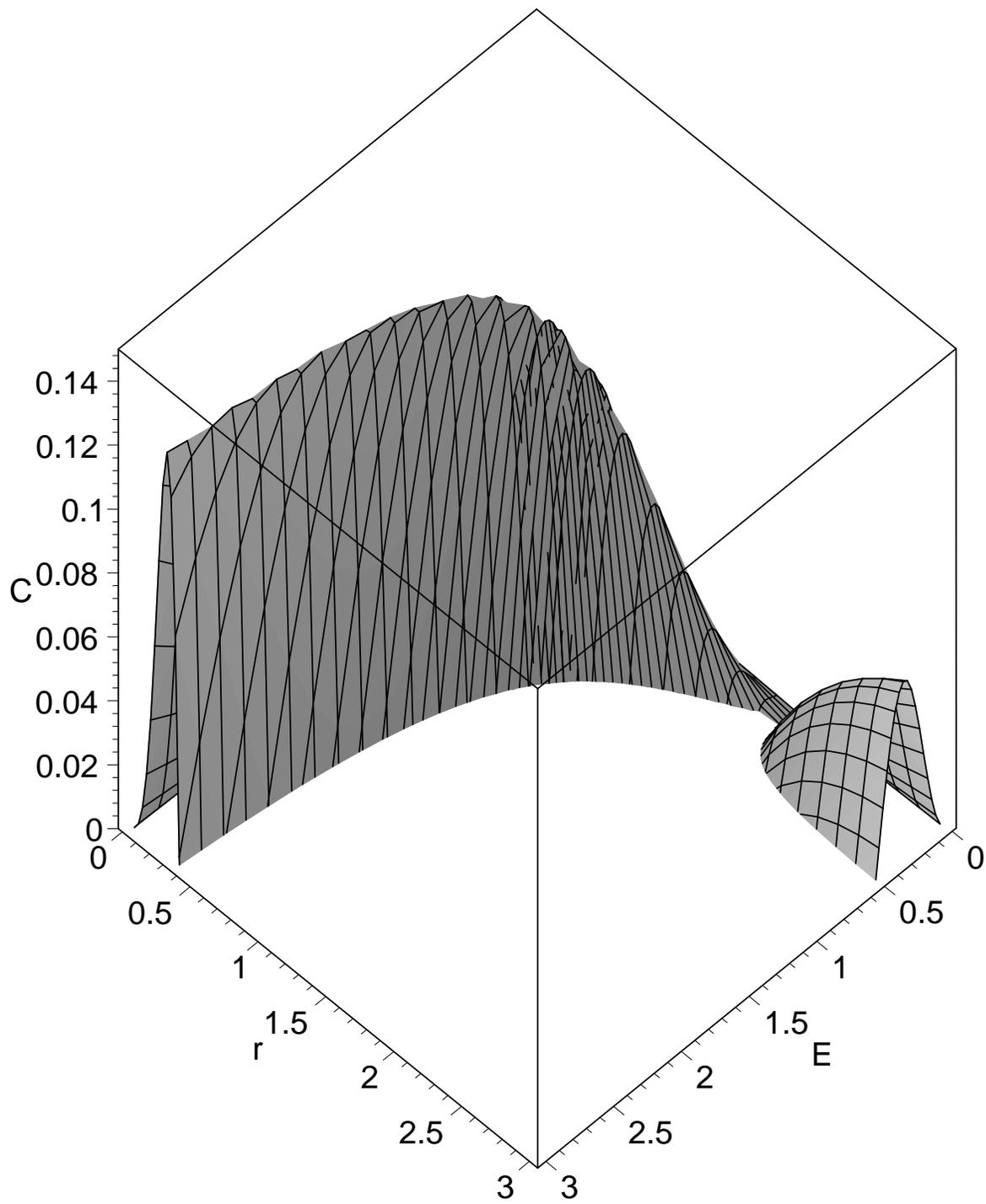
\end{document}